\documentclass[aps,preprint,superscriptaddress,showpacs,preprintnumbers,amsmath,amssymb]
{revtex4}

\input{epsf}

\newcommand{\be}{\begin{equation}}
\newcommand{\ee}{\end{equation}}
\newcommand{\ba}{\begin{eqnarray}}
\newcommand{\ea}{\end{eqnarray}}

\begin{document}

\title{Influence of aperiodic modulations on first-order transitions:
numerical study of the two-dimensional Potts model}

\author{D. Girardi}
\email{girardi@if.uff.br}
\affiliation{National Institute of Science and Technology for Complex System,
Universidade Federal Fluminense,
27.213-350, Rua Des. Ellis Hermydio Figueira, 783 \\  Volta Redonda, RJ, Brazil}

\author{N. S. Branco}
\email{nsbranco@fisica.ufsc.br}
\affiliation{Departamento de Física,
Universidade Federal de Santa Catarina, 
88040-900, Florianópolis, SC, Brazil}

\date{\today}

\begin{abstract}

        We study the Potts model on a rectangular lattice with aperiodic
modulations in its interactions along one direction. 
Numerical results are obtained using the Wolff algorithm and for
many lattice sizes, allowing for a finite-size scaling analyses to be carried out.
Three different self-dual aperiodic sequences are employed, which leads to more precise results,
since the exact critical temperature is known. We analyze two models, with six
and fifteen number of states: both present first-order transitions on their
uniform versions. We show that the Harris-Luck criterion, originally
introduced in the study of continuous transitions, is obeyed also for first-order ones.
Also, we show that the new universality class that emerges for relevant aperiodic
modulations depends on the number of states of the Potts model,
as obtained elsewhere for random disorder, and on the aperiodic sequence. We determine the
occurrence of log-periodic behavior, as expected for models with aperiodic modulated interactions.

\noindent     

\end{abstract}

\maketitle

\newpage

\section{Introduction} \label{sec:introduction}

    Non-uniform systems are ubiquitous in nature. The non-uniformity
may be a consequence of random disorder or of a deterministic construction
of materials from two different atoms, for example.
Experimentally, there are already several techniques of surface growth 
\cite{ross,shchukin,williams} that let one controls the layout of the layers in order to follow, for
example, an aperiodic sequence.
One of the main theoretical issues is to what extent the
introduction of non-uniformities affects the critical behavior of
a system, when compared to its uniform counterpart. 

    For random
quenched disorder and continuous transitions on the uniform model,
this question is partially answered by the Harris criterion \cite{harris}.
According to this criterion, the random model will have the same
critical behavior of the uniform one if the specific heat critical exponent, $\alpha$, 
of the latter is negative. The disorder is said to be irrelevant in this case.
If $\alpha$ is positive, the critical exponents
of the random model are different from the exponents of its uniform counterpart
and the disorder is said to be relevant.
For $\alpha=0$ (the marginal case), logarithm corrections appear
\cite{pena}. A generalization of this criterion is available also for short-range 
\cite{branco5,branco6,branco7} and long-range \cite{branco8,weinrib} correlated disorder.
When the transition in the uniform model is a first-order one, the scenario is qualitatively different:
in two dimensions, even an infinitesimal amount of disorder  changes the nature of the transition to
a continuous one \cite{aizenman,hui,branco9,branco10} while in three dimensions a finite amount of disorder is
necessary to change the order of the transition \cite{hui}.

    We are mainly interested here in the critical behavior of
models with aperiodic modulation of their interaction parameter(s). For the case of continuous
transition on the uniform model,
the Harris-Luck criterion determines when the introduction of a given aperiodic modulation
changes the universality class of a model, compared to its uniform version \cite{luck1}. This
change is determined by the crossover exponent $\Phi$, given by:
\be
          \Phi = 1 + d_a \nu (\omega - 1)
          \label{eq:criterion}
\ee
where $\omega$ is the exponent describing the behavior of geometrical fluctuations of the sequence 
(see below), $d_a$ is the number 
of dimensions upon which the aperiodic sequence acts and $\nu$ is the correlation-length critical exponent 
of the uniform model. When $\Phi>0$ the sequence is relevant 
and when $\Phi<0 $ the sequence is irrelevant. 

     This criterion applies to continuous transitions in the uniform model
but numerical results \cite{berche,chatelain} indicate that it holds true also when the
transition in the original model is a first-order one (if this is the case, the introduction of
aperiodic modulations on the interaction parameters leads to a scenario totally
different from the one for random disorder).
This result was established through a study of the ferromagnetic Potts model \cite{wu} 
with $q=8$ states in Refs. 
\onlinecite{berche} and  \onlinecite{chatelain}; in this model, 
a dynamical variable with $q$ states is assigned to each site on a given lattice and first-neighbor
sites tend to be in the same state (see below).
For low enough $q$, the transition is continuous, while it is first order
for $q$ above a given value $q_c$ (in two-dimensional lattices, $q_c=4$). For $q=8$,
the Harris-Luck criterion is obeyed and the order of the transition changes, for relevant
aperiodic modulations \cite{berche,chatelain}.

    Here, we will address two issues. First, we want to provide an
independent check of the Harris-Luck criterion for first-order transitions.
Second, to test   the possibility that the new universality 
class for relevant modulations depends on the number of states of the Potts model,
as is the case for random disorder \cite{cardy1,jacobsen}. 

   This paper is organized as follows. In the next section, we review some properties of aperiodic
sequences and define the model we use. In Section \ref{sec:MC} we discuss details of the simulation,
while in Sec. \ref{sec:results} we present and discuss our results. Finally, in the last section
we summarize our findings.

\section{Aperiodic sequences and model}\label{sec:aperiodic}
	
	The aperiodic sequences we use in this work are constructed in a  deterministic way, using a
substitution rule on a two-letter alphabet, such that no primitive cell is present.
As an example of a two-letter sequence, $A $ and $ B $, we have: 
     \be
          s(A) \rightarrow AB,\;\;  s(B) \rightarrow BA. \label{eq:sequence}
     \ee
This notation means that we choose an arbitrary initial letter (usually $A$) and, from the above rule,
we build our sequence in the following way. The following
generation of a sequence is obtained  by replacing every letter $A$ and $B$ 
in the previous generation by $AB$ and $BA$, respectively. In this example, we obtain
 $ABBABAAB...$ as our sequence.

Some properties of the sequences can be characterized by
its substitution matrix ${\cal M}$, which is defined as:
     \be \label{eq:matriz}
          {\cal M}=\begin{pmatrix} n_i^{s(i)} & n_i^{s(j)}  
          \\                    n_j^{s(i) }& n_j^{s(j)}\end{pmatrix},
     \ee
where $n_i^{s(j)}$ is the number of letters $i$ that are generated by
applying the rule $s(j)$. Several features of the sequences are determined by
the eigenvalues of ${\cal M}$. The largest eigenvalue ($\lambda_1$) determines the rate of growth of the number of letters ${\cal N}$, such that ${\cal N}\sim\lambda_1^n$, $n\gg1$, where $n$ is the
number of the generations in the construction of the sequence. The second largest eigenvalue ($\lambda_2$) 
determines the {\it wandering  exponent} $\omega$ (see Eq. \ref{eq:criterion}) through:
     \be
          \omega=\frac{\ln |\lambda_2|}{\ln \lambda_1},
     \ee
such that the fluctuation in one of the letters, $g$, is given by \cite{faria}:
     \be
          g\sim {\cal N}^{\omega}.
     \ee

    In order to study the effects of the introduction of aperiodic modulations on models
with first-order transitions, we use the Potts model with $q$
states, which is described by the Hamiltonian:
     \be
          \mathcal{H}=-\sum_{<i,j>}J_{ij} \delta_{\sigma_i\sigma_j}
     \ee
where the sum is made over all first neighbors, $J_{ij}$ is the
coupling constant between sites $ i $ and $ j $, $\sigma_i(=1,2 ... q)$ represents the
state of site $i$, and $\delta_{\sigma_i\sigma_j}$ is the 
Kronecker delta. 
The total magnetization of the Potts model for
a lattice of linear size $L$ is defined as:
     \be
          m_L(T)=\frac{q\rho_{max}-1}{q-1}
     \ee
where $\rho_{max}$ is the density of the most populous state and $T$ is the temperature.
The susceptibility is
obtained as the fluctuation of the magnetization, namely:
   \be
    \chi_L(T)=\beta N(<m^2>-<m>^2), 
    \ee
where $N$ is the total number of sites.\\

In the model we study, the interactions in one direction (horizontal, say) follow an aperiodic
sequence, such that an interaction on a given position assume a value ($J_A$ or $J_B$),
according to the letter ($A$ or $B$, respectively) occupying the same position on the 
aperiodic sequence. In the vertical direction, the interactions have the same value as for the 
following horizontal bonds (see Fig.  \ref{fig:rede} for an example of the sequence
defined in Eq. \ref{eq:sequence}). Thus, all models we study have $d_a = 1$  (see Eq. 
\ref{eq:criterion}).\\

The three sequences used in this work are presented in
Table \ref{table:sequencias}. Two of them are relevant (three-folding, TF, and paper-folding, PF), in case the
Harris-Luck criterion holds true for first-order transitions as well. Therefore, 
they are expected to change the critical behavior when compared to the uniform model. The third
sequence (Thue-Morse, TM) is expected to be irrelevant and should not change the universality class with
respect to its uniform counterpart. These sequences were chosen because they are self-dual, which allows 
for the exact calculation of their critical temperature, through the relation \cite{berche}:
     \be
          (e^{\beta_c}-1)(e^{r\beta_c}-1)=q
     \ee
where $\beta_c=J_A/k_BT_c$, $k_B$ is Boltzmann's constant, $ T_c $ is the critical
temperature, and $ r $ is the ratio between the coupling constants, $ r=J_A/J_B$. Therefore, $r$
describes the amount of aperiodicity of the model: for $r=1$ we regain the uniform model and for
very small (or very large) $r$ we obtain a series of independent one-dimensional-like systems
(we will comment on this issue below).

\begin{table}
  \centering
\begin{tabular}{l*{4}{c}r}
\hline
Sequence         & substitution rule &  $\omega$ & $ p $ & $ pL_{max}$\\
\hline
\hline
Three-Folding & $A\rightarrow ABA$  & $ 0 $ &$ 3 $&$19683$\\
(TF)          & $B\rightarrow ABB$  &            && \\
\hline
Paper-Folding & $AA\rightarrow BAAA$ & $ 0 $ &$ 2 $&$16384$\\
(PF)          & $AB\rightarrow BAAB$ &            &      &\\
              & $BA\rightarrow BBAA$ &            &&\\
              & $BB\rightarrow BBAB$ &            &&\\
\hline

Thue-Morse   & $A\rightarrow AB$ & $ -\infty $ &$ 2 $&$8192$\\
(TM)          & $B\rightarrow BA$   &            && \\
\hline
\end{tabular}
 \caption{Aperiodic sequences, their substitution rules and wandering exponents
$\omega$. $p L_{max}$ is the largest size of the simulated lattice for $q=6$. Note the
abbreviations for the sequences. The quantity $p$ is the rate of growth of the
sequence, after one application of the substitution rule.}\label{table:sequencias}
\end{table}

\section{Details of the simulation}\label{sec:MC}

     The models we treat are invariant under a duality transformation and the exact
critical temperature ($k_B T_c/J_A$) is known \cite{berche, chatelain}. This knowledge allows for a more precise
calculation of critical exponents, using finite-size scaling, but the simulation at $T_c$ also leads to a severe
critical-slowing down if a single-spin algorithm, like Metropolis \cite{newman}, for example, is used. 
Therefore, we resort to
the Wolff algorithm \cite{newman}, which proved to reduce, by a great amount, critical slowing down.  A
rectangular lattice of size $pL \times L $ is used, with periodic boundary conditions in both
directions. The parameter $ p $ varies according to the rate of growth of the aperiodic sequence. As previously
mentioned, the aperiodic interactions are introduced in only one dimension (the one with
linear size $ pL $).  We choose rectangular lattices to be able to introduce larger aperiodic sequences.

     We first checked that the equilibrium values for some thermodynamic quantities were independent of the
initial configuration. As expected, the time to reach equilibrium increased with lattice size but, for the worse cases, were approximately $100$ times the autocorrelation time, $\tau$. 
This quantity was evaluated through the integral of the
autocorrelation functions for the magnetization and the energy and the larger value was considered.  In order to
calculate thermal averages, configurations $2 \tau$ apart (in Wolff steps - one Wolff step corresponds to building and flipping one
island) were taken, which guarantees that the errors may be calculated using the standard deviation,
both for the magnetization and the energy \cite{newman}. All
averages are calculated using a sample of at least $ 10^4 $ independent configurations (where
two configurations are considered independent if they are at least $2 \tau$ apart).

     For the largest lattices, we parallelized our simulation, such that 20 processes run in different CPU's. 
Since the time to reach equilibrium was, for these lattices, approximately
$100$ times the autocorrelation time, this procedure
allows for an economy in computational time. On the other hand, it prevents us from calculating the autocorrelation
time. To overcome this difficulty, we have simulated long time series for the smallest lattices and have
extrapolated the autocorrelation times for the largest ones. See Fig. \ref{fig:extrapolation_tau} for an example of this
procedure for the PF sequence, for $q=6$ and $r=0.1$;  the extrapolation to $L=8,192$, in this case, leads to $\tau=710,000$, in
Wolff steps. If this task was made in only one processor, it would take more than $8$ months to obtain $1,000$
independent configurations.

\section{Results} \label{sec:results}

   Our first goal is to determined the order of the transition and, as a consequence, provide
an independent test of the Harris-Luck criterion. In order to do this, we use the
so-called Lee-Kosterlitz method \cite{lee}, which works as follows.  The quantity
$F_L(E) = - \ln P_L(E)$ is calculated for various linear lattice sizes, $L$, where $P_L(E)$ is 
the probability distribution for energy $E$.  It is expected that, for first-order transitions, $P_L(E)$ will present 
two peaks of the same height in the thermodynamic limit,
one for the disordered phase (high $E$)
and other for the ordered one (low $E$). For continuous transitions, on the other hand, the presence of only
one peak is expected. One then studies the dependence of $\Delta F \equiv \ln \left( P_L(E_p)/P_L(E_b) \right)$ 
with $L$, where $E_p$ and $E_b$ are the values of energy where $P_L(E)$ has a maximum and a minimum,
respectively.  More precisely, a plot of $\Delta F$ as a function of $1/L$ is made,
in order to obtain the tendency of this barrier in the thermodynamic limit ($L \rightarrow \infty$).
If $\Delta F$ goes to zero in this limit, the transition is a continuous one; otherwise, it is a first-order
transition.

      In Fig. \ref{fig:PLdeE} we plot examples of the behavior of $P_L(E)$ for 
what we expect to be continuous ($a$) and first-order
($b$) transitions, for $r=0.9$ and $q=6$, for the $TM$ and $PF$ sequences.
Note that the two-peak structure is present for the $TM$ sequence, while for 
the $PF$ one a crossover effect takes place and only one peak is observed for the largest 
value of $L$. 
  
% \begin{figure}
%\epsfxsize=7.5cm
%\begin{center}
%\leavevmode
%\epsffile{imagens/PLdeE.eps}
%\caption{Plot of $P_L(E)$ (see text) as a function of the energy per spin for
%the $PF$ sequence (a) and for the $TM$ sequence (b), for various values of the linear lattice size
%$L$.} \label{fig:PLdeE} %fig3
%\end{center}
%\end{figure}

   In order to put our results on a more quantitative ground, the behavior of the energy barrier as
a function of $1/L$ is plotted in Fig. \ref{fig:DeltaF}, for a  homogeneous model and for the three
sequences studied here with $r=0.9$ and $q=6$. Note the distinct behavior for the homogeneous 
model and for the
$TM$ sequence, when compared to the $PF$ and $TF$ sequences. While $\Delta F$ increases for $L \rightarrow \infty$ for the former two models, it goes to zero, in the same limit,
for the $PF$ and $TF$ sequences, after a subtle increase for small values of $L$. This indicates that
the transitions is continuous for the $PF$ and $TF$ sequences and first-order for the $TM$ sequence.
This result is consistent with the Harris-Luck criterion. In order to test if this behavior is a consequence
of a value of $r$ close to unity (homogeneous model), we have also performed the same procedure
for $r=0.7$ (inset of Fig. \ref{fig:DeltaF}) for the $TM$ sequence, comparing with the behavior for $r=0.9$:
it is still evident that $\Delta F$ increases with $L$ and no sign of a crossover is seen, for the range
of $L$ studied.

    Note that the Harris-Luck criterion depends on the homogeneous
model only through the value of $\nu$ (see Eq. \ref{eq:criterion}). For models with first-order transitions,
one expects that
$\nu = 1/d$ \cite{fisher}, where $d$ is the dimension of the system. 
Specifically for the Potts model, the value of $\nu$ will not depend on the number
of states $q$, for $q>4$ in two dimensions. So, it is expected that our conclusions will hold true
for other values of $q$ such that the phase transition is a first-order one. Although not shown here,
our results for $q=15$ support this picture.

   Therefore, our calculations suggest that the Harris-Luck criterion is also obeyed when the homogeneous
model presents a first-order transition.  In fact, for $\nu=1/d=1/2$ and $d_a=1$, the crossover exponent
$\Phi$ reads (see Eq. \ref{eq:criterion}):
\begin{equation}
    \Phi = \frac{1}{2} ( \omega + 1 ).  
\end{equation}
 Looking at the values of $\omega$ from Table \ref{table:sequencias}, we see that $\Phi$ is negative 
 (irrelevant sequence), $\Phi=-\infty$, for the $TM$ sequence and positive (relevant sequences),
 $\Phi=1/2$, for the $TF$ and $PF$ sequences. Our findings agree with the results of Refs.
  \onlinecite{berche} and \onlinecite{chatelain}.
  
     We now turn to the question of the new universality classes that emerge when the aperiodic
sequence is a relevant one, which is the case for the $TF$ and $PF$ sequences. We will also
address the possibility that the new universality class depends on the number of states of the
Potts model. In order to calculate the critical exponents, we make simulations at the exact
critical temperature $T_c$ (these two sequences are self-dual, which allows for the exact
calculation of $T_c$). The quantities we calculate are the magnetization $m$ and the susceptibility
$\chi$, which are expected to behave, at $T=T_c$, as:
\begin{equation}
   m \sim L^{-\beta/\nu}, \;\; \chi \sim L^{\gamma/\nu}, 
\end{equation}
where $L$ is the linear size of the lattice and $\nu$ is the critical exponent of the correlation length.
Log-periodic corrections to the above behavior are expected, for aperiodic models, and, as we will
see shortly, are found in our results. We have calculated the two quantities cited above for $q=6$ and $q=15$,
in order to test the dependence of the new critical exponents with the number of states.

     An example of the results we obtain is presented in Fig.  \ref{fig:mag_three}: the log-log graphs
of the magnetization (a) and susceptibility (b) for all three sequences, for 
$r=0.1$ and $q=6$, are depicted. The exponents $ \beta/\nu $ and $\gamma/\nu$ 
are the slope  of the curves $m_L(T_c) \times pL$ and
 $ \chi_L(T_c) \times pL $, respectively. Note the evident crossover behavior for the $TM$ sequence,
 which is an irrelevant one. The slope of the magnetization curve initially follows the behavior of
 the curves for the $PF$ and $TF$ sequences but eventually presents a curvature and the slope tends
 to values closer to zero. Although this is not our stronger evidence for a first-order transition, the
 behavior is consistent with the one for $\Delta F$. The same trend is obtained for the
 susceptibility and a crossover to a different behavior is obtained for the $TM$ sequence.
 
      One possible way to obtain more precise values for the critical exponents is to make
fittings for different ranges of $pL$. More precisely, we initially make fittings for $pL \ge 32$ , which
defines our first estimate. The second estimate is obtained removing the point with the smallest value
of $pL$ used in the first estimate. The procedure is repeated until only three points remain, which
defines our last estimate  \cite{berche,chatelain,picco}. We then plot these values versus
$1/pL_{min}$, where $pL_{min}$ is the smallest value of $pL$ used in a given fitting.The results
for $\beta/\nu$ and $\gamma/\nu$ are depicted in Fig. \ref{fig:exp_three}: it appears that the
values for the $TM$ sequence converge to $\beta/\nu=0$ and $\gamma/\nu=2$, which are
the expected values for a first-order transition \cite{fisher}. For the $PF$ and $TF$ sequences, 
our results are analogous to
the ones in Ref. \onlinecite{berche,chatelain}. However, a detailed look at 
Fig. \ref{fig:exp_three} $b$ and $c$
for small values of $1/pL_{min}$ shows that the convergence to the thermodynamic limit is not
the usual one: oscillations are present for both $\beta/\nu$ and $\gamma/\nu$ and they do not die out
for the largest lattices.
This behavior maybe a sign of log-periodic oscillations,
expected to be present in models with aperiodic modulations \cite{andrade1,andrade2}.

    In order to test this possibility, we have calculated the ratio of the exponents in the following way.
For each three successive points of $m(T_c)$ or $\chi(T_c)$ we fit a straight line and plot
its slope as a function of $1/p\tilde{L}$, where $p\tilde{L}$ is the smallest value of the three
used to make a given fit. For the ``usual'' behavior, the value of these 
slopes should converge to the thermodynamic value of $\beta/\nu$ or $\gamma/\nu$, 
respectively \cite{picco}. In Fig. \ref{fig:fit_three}, we show the results of this procedure for the ratio
$\beta/\nu$, for the $PF$ sequence, $r=0.1$, and $q=6$ and $q=15$.
As clearly seen in this figure, our results show no sign of this
expected convergence. In fact, our data is well fitted by a log-periodic function, in the form: 
\begin{equation}
          \Theta(L) = \Theta + A \cos \left[ B \log(L) + C \right],  \label{eq:log-periodic_function}
\end{equation}
where $\Theta(L)$ stands for the slope of the straight line at a given $L$ and $\Theta$ stands for
$\gamma/\nu$ or $\beta/\nu$, depending on whether we use the data for
$\chi(T_c)$ or $m(T_c)$, respectively.  This shows the expected behavior for systems with
aperiodic modulations \cite{andrade1,andrade2}.

   But note that Eq. \ref{eq:log-periodic_function} has 4 fitting parameters and this requires a
reasonable amount of data points. We can accomplish this for the $PF$ sequence and good fittings are
obtained in this case. However, two issues should be addressed here. First, note that fewer data 
points are available for the $TF$ sequence, in comparison to the $PF$ one (this is due to the larger rate of 
growth of the former sequence). Therefore, one should anticipate that a fitting using 
Eq. \ref{eq:log-periodic_function} might not be satisfactory. We have, therefore, applied a second
procedure to obtain $\gamma/\nu$ and $\beta/\nu$ for the $PF$ sequence: we ignore the log-periodic 
oscillation and fit the data to a  straight line. In this case, only two parameters are necessary and one
hopes that fewer points are needed to provide a good fitting. In Table \ref{table:compare} we compare
$\beta/\nu$ and $\gamma/\nu$ calculated using the two procedures explained above, for the $PF$
sequence and $q=6$ and $15$. One can note that, within error bars (assumed here as one standard
deviation), the values for straight-line fitting and with the use of a log-periodic function 
(Eq. \ref{eq:log-periodic_function}) coincide. This corroborates our strategy for the $TF$ sequence:
since for this sequence a precise fitting using a log-periodic function is not possible, we will rely 
on the straight-line fitting, to obtain $\gamma/\nu$ and $\beta/\nu$.

\begin{table}
  \centering
\begin{tabular}{| r | c | c | c | c |} \hline
 & \multicolumn{2} {| c |}  {$\beta/\nu$} & \multicolumn{2} {| c |}  {$\gamma/\nu$} \\ \hline
  q  & S & LP & S & LP \\ \hline
  6   & 0.469(5)    & 0.470(4)    & 1.032(12) & 1.03(1) \\
  15 & 0.509(16)  & 0.499(10) & 0.93(6)     & 0.93(2) \\ \hline
\end{tabular}
 \caption{Values for the ratios $\beta/\nu$ and $\gamma/\nu$ for the $PF$
 sequence, calculated through a straight-line fitting ($S$) and using a log-periodic
 function ($LP$). $q$ is the number of states. Each  entry above is the average over three
 values of $r$ (see text).}\label{table:compare}
\end{table}

       Also from Table \ref{table:compare}, we notice that the values for
 $\gamma/\nu$ and $\beta/\nu$ depend on the number of estates $q$. This is the same result as
 for random disorder \cite{cardy1,jacobsen,chatelain2}. Note also that although our values for
 $\beta/\nu$ are not equal to those for random disorder \cite{chatelain2}, for a given $q$, the percentage
difference between the values for $q=6$ and $q=15$ are the same for our model and for the model
studied in Ref.\onlinecite{chatelain2}.

   The second issue we would like to address is the presence of crossover effects, for values of $r$
close to $1$ and close to zero. For the former, one expects ``contamination''
from the uniform behavior, with a crossover effect and effective exponents, not representing
the behavior of the aperiodic model. On the other hand, for
$r=0$ the lattice breaks into isolated strips, which are one-dimensional. Therefore, one has
to take into consideration the possibility that a crossover from the one-dimensional model
to the aperiodic behavior may take place. This would be shown as a dependence of the
critical exponents on $r$ \cite{picco}; in fact, from our results obtained for $r$ ranging from $0.001$
to $0.5$, we notice that reasonable stable values for $\beta/\nu$ and $\gamma/\nu$ are obtained
for $r$ between $0.05$ and $0.1$. This means that the aperiodic-model universality class is located
in this interval. In Fig. \ref{fig:expoentes_com_r} we show the behavior of the two ratios mentioned
above as functions of $r$, for the $PF$ sequence and $q=6$ (our results for the $TF$ sequence or
for $q=15$ follow this same behavior). Since we cannot be more precise in locating the value of $r$
which represents the behavior of the aperiodic model, we take our values for $\gamma/\nu$ and
$\beta/\nu$ as the average of the values calculated for $r=0.1$, $0.067$, and $0.05$. Note that
$\gamma/\nu$ increases  for small $r$ and for $r$ close to $1$. This is the expected behavior, since 
for $r=1$ $\gamma/\nu$ is expected to assume the value $2$ in two dimensions \cite{berche,fisher}.
The same is true for $r$ close to zero, since in this limit the one-dimensional character of the model
should lead to an exponential divergence of the susceptibility. For $\beta/\nu$, the decrease in its value
for $r$ close to 1 and zero reflects the fact that in these two limits the transition is first-order, such that
$\beta=0$ \cite{berche,fisher}. Note that our estimate for the location of the aperiodic-model universality
class  coincides with the result for random-disorder Potts-model universality class \cite{chatelain2}.

   Finally, we turn to the $TF$ sequence. As mentioned above, we do not have enough points to make
a reliable fitting with a log-periodic function. Therefore, we resort to straight-line fittings, to obtain
$\beta/\nu$ and $\gamma/\nu$. We have made simulations for many values of $r$ and noticed
that the stable region is between $r=0.05$ and $r=0.1$ (the exact point where the aperiodic-model
universality class is present may vary from one sequence to another and from $q=6$ to $q=15$,
but we do not have enough precision to pinpoint the correct value of $r$). So, we have taken as our
results averages over $r=0.1$, $0.067$, and $0.05$. In Table \ref{table:expoentes_sequencias} we
summarize our results for both sequences and for $q=6$ and $15$, for straight-line fittings.
     
 \begin{table}
  \centering
\begin{tabular}{| r | c | c | c | c | c | c |} \hline
     & \multicolumn{3} {| c |}  {$PF$} & \multicolumn{3} {| c |}  {$TF$} \\ \hline
  q  & $6$ & $8$ & $15$ & $6$ & $8$ & $15$ \\ \hline
  $\beta/\nu$        & 0.469(5)    & 0.49(1)    & 0.509(16) & 0.433(4) & 0.44(2) & 0.47(2) \\ \hline
  $\gamma/\nu$  & 1.032(12)  & 1.01(1)   & 0.93(6)       & 1.13(1)   & 1.09(3) & 1.00(3) \\ \hline
\end{tabular}
 \caption{Values for the ratios $\beta/\nu$ and $\gamma/\nu$ for $PF$ and $TF$
 sequences, calculated through a straight-line fitting. $q$ is the number of states. Each  
 entry above is the average over three
 values of $r$ (see text). Numbers between parenthesis represent the error
 in the (two) last digit(s). Results for $q=8$ are from Ref. \onlinecite{chatelain},
 for $r=10$ (equivalent to our $r=0.1$)}\label{table:expoentes_sequencias}
\end{table}    
     
    Our results strongly suggest that the new universality class that emerges when aperiodic modulations
are introduced in the Potts model depends on the number of states $q$ and on the aperiodic sequence.
This latter conclusion is supported by the calculations in Ref. \onlinecite{chatelain}. Together with
the results of Ref. \onlinecite{berche} and  \onlinecite{chatelain}, we can establish that $\beta/\nu$
($\gamma/\nu$) increases (decreases) with $q$, but we are not able to propose a dependence law.
Finally, we would like to mention that our results, shown in Table \ref{table:expoentes_sequencias},
for both sequences, and both values of $q$, satisfy the equality $2 \beta/\nu + \gamma/\nu = 2$,
within error bars.
  
\section{Conclusion} \label{sec:conclusion}

   We have performed a numerical simulation of Potts models on the square lattice with aperiodic
modulations of the interaction parameter in one direction. The Wolff algorithm was used and three
different aperiodic sequences were studied, for $q=6$ and $15$ number of states. For these
cases, the transition on the uniform model is a first-order one. Using the Lee-Kosterlitz
method, we verified that the Thue-Morse sequence does not change the universality class of the 
transitions, while the Paper Folding and Three Folding sequences turn the transition into a
continuous one. These results are in accordance to the Harris-Luck criterion and confirm
and extend the results of Refs. \onlinecite{berche} and \onlinecite{chatelain2}. Therefore, we
expect that this criterion applies to first-order transitions, although it was introduced in the
context of models with continuous transitions.

   For the two sequences which are relevant, we calculate the universality class, simulating the systems
on theirs exact critical temperature. Using finite-size scaling arguments, we were able to show that
this new universality class depends on the number of states $q$ and on the aperiodic sequence. This latter
result is in accordance with earlier calculations \cite{berche,chatelain2}.

  We have also made simulations to calculate the exponent of the correlation length, $\nu$, studying
quantities such as the fourth-order cumulant and logarithmic derivatives of the magnetization
\cite{chen}. However, strong oscillations, already seen in  Refs. \onlinecite{berche} and  
\onlinecite{chatelain2}, prevented us from getting precise results.

    In order to calculate $\nu$, obtain more precise values for the value of $r$ which represents
the critical behavior of the aperiodic model, and go to larger values of $q$, we are now using
a transfer matrix matrix approach. Mapping the Potts model onto a Whitney polynomial \cite{blote}, the
number of states $q$ enter as a parameter and the matrix involved grow much slower than
$q^N$, where $N$ is the number of rows of the finite system. This allows for more precise values, for
large $q$, than using Monte Carlo simulation.

\begin{acknowledgments}
The authors would like to thank the Brazilian agencies
FAPESC, CNPq, and CAPES for partial financial support. The referees of a previous version of this paper
are also thanked for their valuable comments.
\end{acknowledgments}

\bibliography{references}

\newpage

\begin{figure}%Figure1%
%\epsfxsize=7.5cm
%\begin{center}
%\leavevmode
%\includegraphics[width=0.45\textwidth]{imagens/rede}
%\epsffile{imagens/rede.eps}
\caption{Interactions with strength $J_A$ and $J_B$ are represented by traced and continuous lines,
respectively. Note that the aperiodicity is introduced into the system in only one direction (horizontal, in this
figure). In the vertical direction, all interactions have the same value as for the following horizontal
bonds. The aperiodic sequence used is the one defined in Eq. \ref{eq:sequence}.}
\label{fig:rede} %fig1
%\end{center}
\end{figure}

\begin{figure}%Figure2%
%\epsfxsize=7.5cm
%\begin{center}
%\leavevmode
%\includegraphics[width=0.45\textwidth]{imagens/extrapolation_tau}
%\epsffile{imagens/extrapolation_tau.eps}
\caption{Log-log graph of the autocorrelation time $\tau$ as a function of the linear size of the lattice, $L$,
for the PF sequence with $q=6$ and $r=0.1$.}
\label{fig:extrapolation_tau} %fig2
%\end{center}
\end{figure}

\begin{figure}%Figure3%
%      \vspace{0cm}
%     \centering
%    \subfigure{
%         \label{fig:PLdeE(a)}
%          \includegraphics[width=0.45\textwidth]{imagens/PL(E)a}}
%     \hspace{0.2cm}
%     \subfigure{
%          \label{fig:PLdeE(b)}
%          \includegraphics[width=0.45\textwidth]{imagens/PL(E)b}}\\
     %\vspace{.3in}
\caption{Plot of $P_L(E)$ (see text) as a function of the energy per spin for
the $PF$ sequence (a) and for the $TM$ sequence (b), for various values of the linear lattice size
$L$, for $r=0.9$ and $q=6$, at $T_c(L)$. }
\label{fig:PLdeE} %fig3
\end{figure}

 \begin{figure}%Figure4%
%\epsfxsize=7.5cm
%\begin{center}
%\leavevmode
%\includegraphics[width=0.45\textwidth]{imagens/DeltaF}
%\epsffile{imagens/DeltaF.eps}
\caption{The energy barrier $\Delta F$ as a function of $1/L$ for the homogeneous model and
the $TM$, $PF$, and $TF$ sequences, for $r=0.9$ and $q=6$ (main graph). In the inset, a 
comparison of the behavior of
the energy barrier is made for the $TM$ sequence and $r=0.9$ and $r=0.7$.}
\label{fig:DeltaF} %fig4
%\end{center}
\end{figure}

\begin{figure}%Figure5%
%\epsfxsize=7.5cm
%\begin{center}
%\leavevmode
%\includegraphics[width=0.45\textwidth]{imagens/mag_susc_todos}
%\epsffile{imagens/mag_susc_todos.eps}
\caption{Log-log graph of the magnetization (a) and susceptibility (b) versus $pL $ for  the
three aperiodic sequences we study, for $q=6$ and $ r =0.1$. Full lines correspond to power-law
fittings and dotted lines are just guides to the eye. Error bars are smaller than the points.}
\label{fig:mag_three} %fig5
%\end{center}
\end{figure}

\begin{figure}%Figure6%
%\epsfxsize=7.5cm
%\begin{center}
%\leavevmode
%\includegraphics[width=0.45\textwidth]{imagens/lmin_expoentes}
%\epsffile{imagens/lmin_expoentes.eps}
\caption{Semi-log graph of the exponents versus $1/pL_{min} $ for the
three-folding sequence, for $r = 0.1$, $q = 6$ and for the three sequences studied.}
\label{fig:exp_three} %fig6
%\end{center}
\end{figure}

\begin{figure}%Figure7%
%\epsfxsize=7.5cm
%\begin{center}
%\leavevmode
%\includegraphics[width=0.45\textwidth]{imagens/fit_pf}
%\epsffile{imagens/fit_pf.eps}
\caption{Graphs of $\beta/\nu$ versus $1/p\tilde{L}$ (see text) for the $PF$ sequence, $r=0.1$, 
and $q=6$ (a) and $q=15$ (b)}
\label{fig:fit_three} %fig7
%\end{center}
\end{figure}

 \begin{figure}%Figure8%
%\epsfxsize=7.5cm
%\begin{center}
%\leavevmode
%\includegraphics[width=0.45\textwidth]{imagens/Expoentes_com_r}
%\epsffile{imagens/Expoentes_com_r.eps}
\caption{Graphs of the ratios $\gamma/nu$ (main graph) and $\beta/\nu$ (inset) 
as a function of $log(r)$ for the $PF$ sequence and $q=6$.}
\label{fig:expoentes_com_r} %fig8
%\end{center}
\end{figure} 

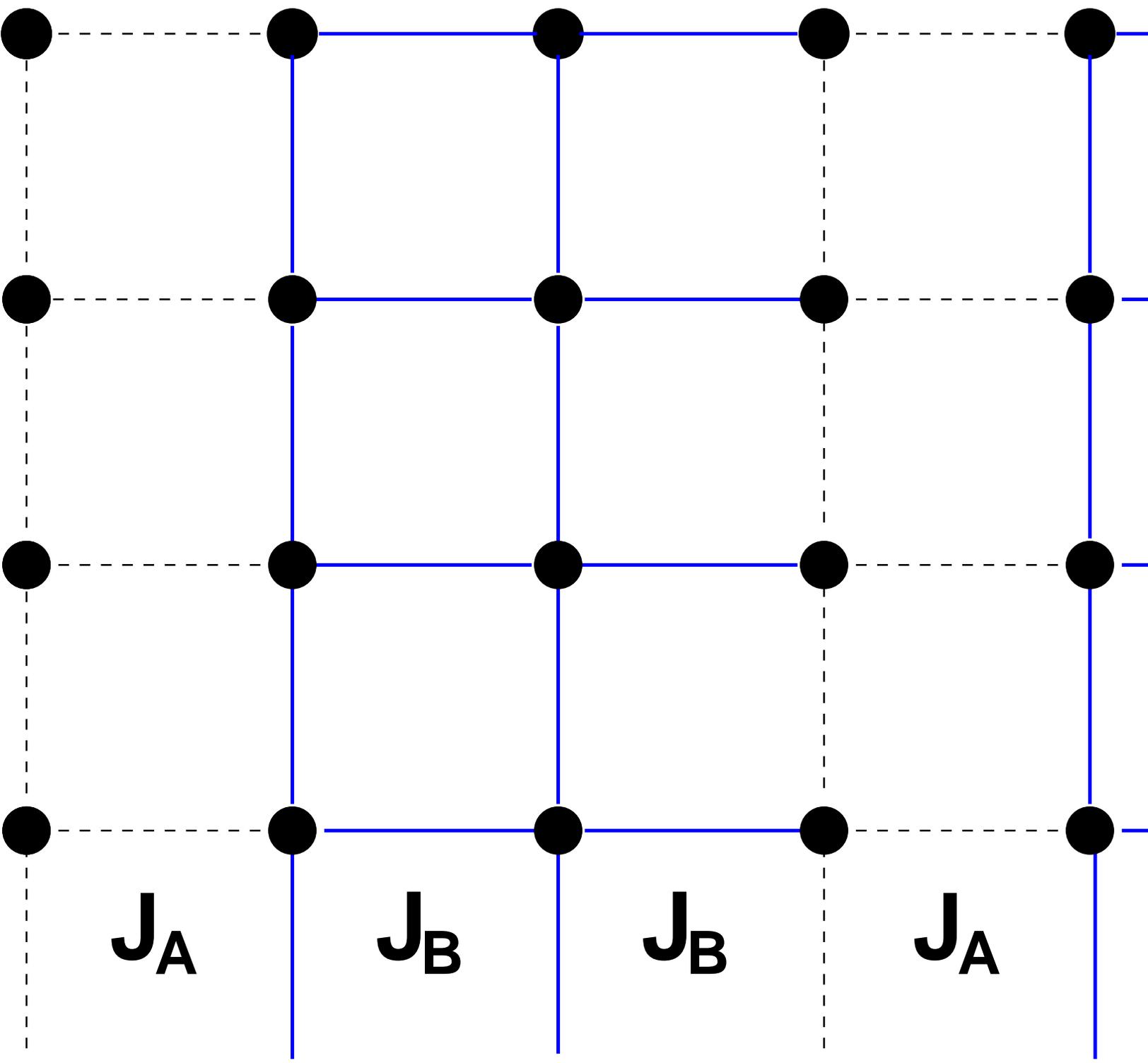
\end{document}